\pdfoutput=1
\documentclass[12pt]{article}

\font\grande=cmr9.5 scaled \magstep4
\font\medio=cmr9.5 scaled \magstep2
\outer\def\beginsection#1\par{\medbreak\bigskip
      \message{#1}\leftline{\bf#1}\nobreak\medskip
\vskip-\parskip
      \noindent}
\usepackage{graphicx} 
\begin{document}
\bibliographystyle {unsrt}

\titlepage

\begin{flushright}
%%%%%%
\end{flushright}

\vspace{15mm}
\begin{center}
{\grande Spectator electric fields, de Sitter space-time}\\
\vspace{5mm}
{\grande and the Schwinger effect}\\
\vspace{15mm}
 Massimo Giovannini 
 \footnote{Electronic address: massimo.giovannini@cern.ch} \\
\vspace{0.5cm}
{{\sl Department of Physics, Theory Division, CERN, 1211 Geneva 23, Switzerland }}\\
\vspace{1cm}
{{\sl INFN, Section of Milan-Bicocca, 20126 Milan, Italy}}
\vspace*{1cm}

\end{center}

\centerline{\medio  Abstract}
\vskip 0.5cm
During a de Sitter stage of expansion the spectator fields of different spin are 
constrained by the critical density bound and by further requirements determined by their  
specific physical nature. The evolution of spectator electric fields in
 conformally flat background geometries is occasionally concocted  
 by postulating the existence of ad hoc currents but this apparently innocuous
 trick violates the second law of thermodynamics.  
Such a problem occurs, in particular, for those configurations 
(customarily employed for the analysis of the Schwinger 
effect in four-dimensional de Sitter backgrounds) leading to an electric energy density which 
is practically unaffected by the expansion of the underlying geometry. 
The obtained results are compared with more mundane situations 
where Joule heating develops in the early stages 
of a quasi-de Sitter phase. 
\noindent

\vspace{5mm}

\vfill
\newpage
Spectator fields are a key element of quantum theories in 
curved background geometries  (see \cite{one} for two classic
monographs on this theme). Their energy density is by definition negligible 
in comparison with the critical energy density of the background: in the opposite case
they would not spectate the dynamics but instead modify the evolution of the underlying geometry. 
In the expanding de Sitter space-time and in some of its inflationary extensions
sub-critical fields induce a number of diverse physical effects that
further constrain their evolution. Of particular interest have been, in recent years, 
the spectator gauge fields appearing in different frameworks ranging 
from magnetogenesis \cite{two} to anisotropic inflation \cite{three}. 
In the latter context the spectator fields may become dominant and 
lead to an explicit breaking of the spatial isotropy by so causing computable 
(and phenomenologically constrained) anisotropic corrections to the scalar and tensor power spectra \cite{four}. 
Spectator (electric) fields also appear in the analysis of the Schwinger effect in 
quasi-de Sitter space-time \cite{five}: in this context the constancy 
of the energy density is achieved by considering a class 
of homogeneous (electric) field configurations sustained by an appropriate current. 
This class of spectator  fields, superficially allowed in a de Sitter (or quasi-de Sitter) space-time 
by analogy with the Minkowskian situation, causes a violation 
of the second law of thermodynamics.

The contravariant components of the gauge 
field strength in conformally flat spacetimes of Friedmann-Robertson-Walker type\footnote{In the first 
part of the analysis the background metric is simply given by $g_{\mu\nu} = a^2(\tau) \eta_{\mu\nu}$; 
$\eta_{\mu\nu} = \mathrm{diag}(1, \, -1,\, -1,\, -1)$ is the Minkowski metric, $a(\tau)$ is the 
scale factor and $\tau$ denotes the conformal time coordinate. The derivation will be extended to the generally covariant 
case in the second part of the paper.} are notoriously given by $Y^{i0} = e^{i}/a^2$ and 
$Y^{ij} = - \epsilon^{ijk} b_{k}/a^2$ so that the explicit evolution of 
the comoving electric (i.e. $\vec{E} = a^2 \vec{e}$) and magnetic (i.e. $\vec{B} = a^2 \vec{b}$) 
fields can be expressed as:
\begin{eqnarray}
&& \vec{E}^{\prime} + 4 \pi \vec{J} = \vec{\nabla} \times \vec{B}, \qquad \vec{B}^{\prime} = -  \vec{\nabla} \times \vec{E},
\label{2one}\\
&& \vec{\nabla} \cdot \vec{E} = 4 \pi \rho_{q}, \qquad  \vec{\nabla} \cdot \vec{B} =0,
\label{2three}
\end{eqnarray}
where the prime denotes a derivation with respect to the conformal time coordinate $\tau$;  
the charge density and the comoving current are, respectively, 
$\rho_{q}$ and $\vec{J}$.  Because of Weyl invariance, the well known form of Eqs. (\ref{2one})--(\ref{2three})  
matches exactly the standard Maxwell's equations (see e.g.  the first paper of Ref. \cite{two})
but this analogy is misleading since the components of the corresponding canonical energy-momentum 
tensor are not Weyl invariant. In particular, the electric and magnetic energy densities are diluted as:
\begin{equation}
\rho_{B}(\tau, \vec{x}) = \frac{B^2(\tau,\vec{x})}{8\pi a^4(\tau)}, 
\qquad \rho_{E}(\tau,\vec{x}) = \frac{E^2(\tau,\vec{x})}{8\pi a^4(\tau)}.
\label{2four}
\end{equation}
While in Minkowski space the time-independent electromagnetic fields generally lead to a 
constant energy density, Eqs. (\ref{2one}), (\ref{2three}) and (\ref{2four}) suggest the opposite:
the electromagnetic fields in a conformally flat space-time of Friedmann-Roberston-Walker type
may very well be comovingly constant (i.e. $\vec{E}^{\prime} = \vec{B}^{\prime} =0$ and $\vec{\nabla} \times \vec{E} = 
\vec{\nabla}\times \vec{B}=0$) without leading to a time-independent electromagnetic energy density. 
Indeed, the simplest gauge configurations compatible with Eqs. (\ref{2one})--(\ref{2three}) are the ones
where {\it i)} the fields are comovingly constant (i.e. $\vec{E}(\tau, \vec{x}) = \vec{E}_{c}$ and 
$\vec{B}(\tau, \vec{x}) = \vec{B}_{c}$),  {\it ii)} the system is globally neutral (i.e. $\rho_{q} =0$) and 
{\it iii)} the total current is absent  (i.e. $\vec{J}=0$). This is the free adiabatic 
evolution of the gauge fields: the sources are absent and the electromagnetic
 energy densities redshift as $a^{-4}$ becoming quickly negligible as the Universe expand (see Eq. (\ref{2four})). 
 This conclusion holds, a fortiori, in the case of a quasi de Sitter stage of inflationary expansion provided 
 $\vec{E}$ and $\vec{B}$ are bona fide spectator gauge fields. By pursuing the same logic 
 the solutions of free Maxwell's equations in Minkowskian space-time can be lifted to the case of 
 a conformally flat background geometry with the important proviso that their energy-momentum tensor is not Weyl 
invariant. Consequently, if we set the Cauchy data at some initial time $\tau_{i}$ by requiring the
 the electric and magnetic energy densities be smaller than the critical energy 
 density\footnote{We shall be using the standard notations implying that the Hubble rate  
 $H = a {\mathcal H}$ where ${\mathcal H} = a^{\prime}/a$; moreover 
$\overline{M}_{P} = 1/\sqrt{8\pi G}$.}(i.e.  $\rho_{E}(\tau_{i}, \vec{x}) < 3 H^2 \overline{M}_{\mathrm{P}}^2$ 
and $\rho_{B}(\tau_{i}, \vec{x}) < 3 H^2 \overline{M}_{\mathrm{P}}^2$), in the free adiabatic case 
the role of gauge fields will be progressively immaterial as the Universe inflates. 

The energy densities does not necessarily 
scale as in the free adiabatic case when the sources are switched on.
For instance in the case of a globally neutral and conducting plasma 
\begin{equation}
\vec{J}= \sigma \vec{E}, \qquad \sigma = a \sigma_{c}, \qquad \rho_{q} =0,
\label{4zero}
\end{equation} 
where $\sigma$ denotes the comoving conductivity while $\sigma_{c}$ is commonly referred to as the physical conductivity. 
If and when the plasma is relativistic $\sigma_{c}$ is proportional to the physical temperature and 
the comoving conductivity is constant as a function of $\tau$. The situation described by Eq. (\ref{4zero}) arises during the early 
stages of a quasi-de Sitter stage of expansion 
as a possible remnant of a pre-inflationary epoch (see e.g. the second and third papers of Ref. \cite{two}).
Yet a different possibility is that the electric fields 
be constant in space-time so that an {\em inhomogeneous} solution of Eqs. (\ref{2one}) and (\ref{2three}) is:
\begin{equation}
\vec{\nabla} \times \vec{B} = 4 \pi \vec{J}, \qquad \vec{\nabla} \cdot \vec{B} = \vec{\nabla} \cdot \vec{J} = 0, \qquad 
\rho_{B}(\tau,\vec{x}) = \frac{B^2(\vec{x})}{a^4(\tau)}, \qquad \rho_{E}(\tau) = \frac{E_{c}^2}{a^4(\tau)},
\label{4one}
\end{equation}
where $E_{c}$ is a constant field and $\vec{B}^{\,\prime}=0$. Conversely when the magnetic fields are constant 
in space-time, the {\em fully homogeneous} electric fields must 
obey\footnote{Note that in Eq. (\ref{4two}) we included $B_{c}$ corresponding to a 
homogeneous magnetic background; however this is not necessary for the present discussion so 
that, in practice, we shall be focussing on the case $B_{c} =0$.}: 
\begin{equation}
\vec{E}^{\prime} + 4 \pi \vec{J} =0, \qquad \vec{\nabla} \times \vec{B} = \vec{\nabla} \times \vec{E}= 0,\qquad 
 \rho_{E}(\tau) = \frac{E^2(\tau)}{a^4(\tau)}, \qquad \rho_{B}(\tau) = \frac{B_{c}^2}{a^4(\tau)}.
\label{4two}
\end{equation}
The first relation of Eq. (\ref{4two}) suggests that {\em any} functional form of the homogeneous 
electric field can be safely reproduced by postulating the existence of an appropriate comoving current.
For the sake of simplicity the evolution of the electric field can be parametrized as
$\vec{E}(\tau) = E_{0} a^{\lambda}(\tau) \hat{n}$ where $E_{0}$ is a constant, $\hat{n}$ 
is a unit vector and $\lambda$ is the parameter expressing the evolution of the electric field in units 
of the underlying rate of evolution. Recalling Eq. (\ref{2four}) the electric energy density will then scale
as $\rho_{E} \propto E_{0}^2 a^{2 \lambda -4}$.  Assuming that initially $E_{0}^2/(H^2 \overline{M}_{P}^2) \ll 1$ 
the critical density bound will always be satisfied at a later time provided $\lambda \leq 2$. 
 When  $\lambda = 2$ the electric field is time-dependent but the electric energy density is a space-time constant:
this is exactly the case of Ref. \cite{five} where, more often than not,  the contribution of the magnetic field 
has been neglected by setting, in our notations, $B_{c} =0$. All in all Eqs. (\ref{2four}) and (\ref{4two}) 
demonstrate that when the sources do not vanish in Eqs. (\ref{2one}) and (\ref{2three}) 
the corresponding energy density may be less suppressed than in the free adiabatic 
case or even become a space-time constant. 

While the chain of arguments presented so far is rarely spelled out in concrete analyses,
it is anyway artificial unless the dynamics of the charged species is appropriately discussed. 
In even simpler words: what about the sources leading to the wanted current? 
Are they physically sound? When $\lambda \leq 2$ the comoving charge density 
must vanish while, according to Eq. (\ref{4two}), the comoving current is  given by:
\begin{equation}
\vec{J} = - \frac{\lambda {\mathcal H} }{4 \pi} \,a^{\lambda}\, E_{0} \,\hat{n} \equiv - \frac{\lambda\, 
{\mathcal H}}{4\pi} \vec{E}, \qquad \rho_{q} =0, \qquad {\mathcal H} = a H = \dot{a}.
\label{5one}
\end{equation}
A swift comparison of Eqs. (\ref{5one}) and (\ref{4zero}) suggests the definition of an effective conductivity 
$\sigma^{(\mathrm{eff})} = - \lambda {\mathcal H}/(4\pi)$ which is {\em negative} when the Universe {\em expands}
(i.e. $\dot{a} >0$) and {\em positive} when the Universe {\em contracts}\footnote{Even if this 
analogy will not be exploited in the subsequent considerations, it is intuitively useful 
to appreciate that, in the present case, the expansion rate determines the sign of the effective 
conductivity.}.  This intuitive analogy already suggests that the current $\vec{J}$ cannot be virtual 
and should rather arise from the the charge carriers entering the total (covariantly conserved) 
energy-momentum tensor whose explicit form can be written as:
\begin{equation}
T^{\mu\nu}_{\mathrm{tot}} = T^{\mu\nu}_{\Lambda} + T^{\mu\nu}_{(+)} +  T^{\mu\nu}_{(-)} + T^{\mu\nu}_{\mathrm{gauge}},
\label{5two}
\end{equation}
where $T^{\mu\nu}_{\Lambda} = - p_{\Lambda} g^{\mu\nu}$ denotes the contribution 
of the cosmological constant and, as usual, $p_{\Lambda} = - \rho_{\Lambda}$.
In Eq. (\ref{5two}) $T^{\mu\nu}_{\mathrm{gauge}}$ and  $T^{\mu\nu}_{(\pm)}$ are, respectively, 
the energy-momentum tensors of the gauge field and of the charged species:
\begin{equation}
T^{\mu\nu}_{\mathrm{gauge}} = \frac{1}{4 \pi} \biggl[ - Y^{\mu\alpha} Y^{\nu}_{\,\,\,\,\,\, \alpha} 
+ \frac{1}{4} g^{\mu\nu} Y_{\alpha\beta} Y^{\alpha\beta} \biggr], \qquad 
T^{\mu\nu}_{(\pm)} = (p_{\pm} + \rho_{\pm}) u^{\mu}_{(\pm)} u^{\nu}_{(\pm)} - p_{\pm} g^{\mu\nu}.
 \label{5twoA} 
 \end{equation}
It is well known that the compatibility with the covariant form of Maxwell equations
\begin{equation}
\nabla_{\mu} Y^{\mu\nu} = 4 \pi j^{\nu}, \qquad  \nabla_{\mu} \widetilde{Y}^{\mu\nu} =0, 
\qquad j^{\nu} =  j^{\nu}_{(+)} +  j^{\nu}_{(-)},
\label{5eightA}
\end{equation}
determines the evolution of the charged components of the energy-momentum tensor;  more specifically we have
\begin{equation}
\nabla_{\mu}  T^{\mu\nu}_{\mathrm{gauge}} = - \biggl[ j^{\alpha}_{(+)} + j^{\alpha}_{(-)}\biggr] Y^{\nu}_{\,\,\,\,\,\alpha},\qquad 
\nabla_{\mu}  T^{\mu\nu}_{(\pm)} = j^{\alpha}_{(\pm)} Y^{\nu}_{\,\,\,\,\,\,\alpha}.
\label{5four}
\end{equation}
The (covariantly conserved) four-currents  appearing in Eqs. (\ref{5eightA})--(\ref{5four}) are, by definition,  
$ j^{\alpha}_{(\pm)} = \pm \,q \,\widetilde{n}_{\pm}\, u^{\alpha}_{(\pm)}$ where $\widetilde{n}_{\pm}$ denote the physical concentration 
while $n_{\pm} = a^3\, \widetilde{n}_{\pm}$  are the corresponding comoving concentration. The total comoving charge $\rho_{q} = q(n_{+} - n_{-})$ vanishes provided $n_{\pm}$ coincide with a common value (be it $n_{0}$). The total comoving 
current of Eq. (\ref{2one}) is $\vec{J} = q \,[n_{+} \vec{v}_{+} - n_{-} \vec{v}_{-}]$
where the four-velocities appearing in  $ j^{\nu}_{(\pm)}$ have been expressed as 
 $u^{i}_{(\pm)} = u_{(\pm)}^{0} \,v^{i}_{\pm}$ by recalling that $u_{(+)}^{0} = u_{(-)}^{0} = 1/a$. 
 Thanks to the global neutrality of the system the total current is simply given by 
 $\vec{J} = q \,n_{0}\, (\vec{v}_{+} - \vec{v}_{-})$. Needless to say that in the case 
 $g_{\mu\nu} = a^2(\tau) \,\eta_{\mu\nu}$,  Eq. (\ref{5eightA}) leads exactly to Eqs. (\ref{2one}) and 
(\ref{2three}). 

Using the explicit expressions of the four-currents Eq. (\ref{5four}) becomes, in a conformally flat geometry of Friedmann-Robertson-Walker type, 
\begin{equation}
\rho_{\pm}^{\prime} + 3 {\mathcal H} (\rho_{\pm} + p_{\pm}) = \pm \frac{q\, n_{0}\, (\vec{v}_{\pm}\cdot \vec{E})}{a^4}.
\label{5five}
\end{equation}
The equations for the positively and negatively 
charged species can be summed up\footnote{The global flow is defined as $(\rho + p) \vec{v} =  (p_{+} + \rho_{+}) \vec{v}_{+} + (p_{-} + \rho_{-}) \vec{v}_{-}$ where $\vec{v}$ is the total velocity which can be interpreted as the centre of mass velocity 
of the charge carriers in the case of massive species. While $\vec{v}$ is related to the sum of the peculiar velocities, the comoving current 
$\vec{J}$ is given by their difference.} by introducing the 
total energy density (i.e. $\rho= \rho_{+} + \rho_{-}$) and  pressure (i.e. $p = p_{+} + p_{-}$):
\begin{equation}
 \rho^{\prime} +  3 {\mathcal H} (\rho + p) = \frac{\vec{J}\cdot \vec{E}}{a^4}, \qquad \vec{J} = q n_{0} (\vec{v}_{+} - \vec{v}_{-}).
\label{5six}
\end{equation}
Since the total current must now be given by the solution of Eq. (\ref{5one}), Eq. (\ref{5six}) becomes:
\begin{equation}
\rho^{\prime} + 3 {\mathcal H} (\rho + p) = - \frac{\lambda \,{\mathcal H}}{4\pi} \, E_{0}^2 \, a^{2 \lambda -4},
\label{5sixa}
\end{equation}
which can be explicitly integrated  for any value of the parameter $\lambda$. However, for the sake of simplicity we shall posit $\lambda= 2$ (where the electric energy density turns out to the constant) so that the solution of Eq. (\ref{5six}) becomes, in this case,
\begin{equation} 
\rho(\tau) = \biggl[ \rho(\tau_{i}) + \frac{E_{0}^2}{6 \pi (w+1)} \biggr] \biggl(\frac{a_{i}}{a}\biggr)^{3(w +1)} - 
\frac{E_{0}^2}{6 \pi ( w+1)},
\label{5nine}
\end{equation}
where $w = p/\rho$ is the constant barotropic index (i.e. $0\leq w \leq 1$) and $\rho(\tau_{i})$ denotes the energy density 
of the charge carriers at the onset of the dynamical evolution, i.e. for an initial time $\tau_{i}$ which might even coincide with the onset of inflation. Equation (\ref{5nine}) implies that as long as $a \gg a_{i}$ and $w$ is positive semidefinite, $\rho$ violates asymptotically the weak energy condition. This occurrence is prevented provided the evolution lasts for a limited 
number of efolds, i.e. 
\begin{equation}
\biggl(\frac{a_{*}}{a_{i}}\biggr) < \biggl[ \frac{6 \pi (w+1) \rho(\tau_{i})}{E_{0}^2} +1\biggr]^{\frac{1}{3(w+1)}}, \qquad E_{0}^2 \ll  \rho(\tau_{i})
\ll 3 H^2 \overline{M}_{P}^2.
\label{5ten}
\end{equation}
To fulfil Eq. (\ref{5ten}) we must imperatively  demand that the 
electric energy density be smaller in comparison with the (initial) energy density 
of the charge carriers, i.e. $E_{0}^2 \ll \rho(\tau_{i})$.
By recalling that the entropy density is $ \varsigma = (\rho + p)/T$, Eq. (\ref{5six}) can be expressed as
\begin{equation}
\varsigma^{\prime} + 3 {\mathcal H} \varsigma = - \frac{4 {\mathcal H}}{T} \rho_{E}, 
\label{5tenA}
\end{equation}
where  $\rho_{E}$ is the electric energy density which is constant in the case $\lambda =2$. Equation (\ref{5tenA}) implies that the entropy four-vector is positive when the background contracts (i.e. ${\mathcal H}<0$) 
and it is negative when the background expands (i.e. ${\mathcal H}>0$). Conversely in the Ohmic case 
the second law of thermodynamics corresponding to Eq. (\ref{5tenA}) stipulates  that 
$\varsigma^{\prime} + 3 {\mathcal H} \varsigma =  8 \pi \sigma \rho_{E}/T$
and it is always obeyed in spite of the sign of the expansion rate. 

Within a fully covariant approach the gauge field strengths can be decomposed in terms of their covariant electric and magnetic components\footnote{The very notion of relativistic electric or magnetic fields might seem self-contradictory since the electric and magnetic fields are non-relativistic concepts that must be replaced, in a Lorentz covariant formulation, by the appropriate field strength tensor $Y_{\mu\nu}$ (and by its dual $Y^{\alpha\beta}= \epsilon^{\alpha\beta\mu\nu} Y_{\mu\nu}/(2 \sqrt{-g})$). However, if there exist a family of four-dimensional observers moving with four-velocity $u_{\mu}$ the relativistic dynamics can be phrased in terms of the 
${\mathcal E}_{\mu}$ and ${\mathcal B}_{\mu}$ introduced in Eq. (\ref{eqLL}).} as argued 
long ago by Lichnerowicz in the discussion of hydromagnetic nonlinearities \cite{six}:
\begin{equation}
Y_{\mu\nu} = {\mathcal E}_{[\mu}\, u_{\nu]}\, + \, E_{\mu\nu\rho\sigma}\, u^{\rho} {\mathcal B}^{\sigma},\qquad 
\widetilde{Y}^{\mu\nu} = {\mathcal B}^{[\mu}\, u^{\nu]}\, + \, E^{\mu\nu\rho\sigma}\, {\mathcal E}_{\rho} u_{\sigma},
\label{eqLL}
\end{equation}
where  $ {\mathcal E}_{[\mu}\, u_{\nu]} = {\mathcal E}_{\mu} u_{\nu} - {\mathcal E}_{\nu} u_{\mu}$ (and similarly for ${\mathcal B}^{[\mu}\, u^{\nu]}$); note that ${\mathcal E}^{\mu} = Y^{\mu\nu}\, u_{\nu}$ and 
${\mathcal B}^{\mu} = \widetilde{Y}^{\mu\nu}\, u_{\nu}$  generalize, in some sense, the electric and the magnetic 
components to the relativistic regime. The covariant evolution of the individual energy momentum tensors
\begin{equation}
\nabla_{\mu} T^{\mu\nu} = Y^{\nu}_{\,\,\,\,\,\,\alpha} \,\,j^{\alpha}, \qquad \nabla_{\mu} T_{\mathrm{gauge}}^{\mu\nu} = -Y^{\nu}_{\,\,\,\,\,\,\,\alpha}\,\, j^{\alpha}, 
\label{VIO1}
\end{equation}
implies that the contribution of the total current cancels from the sum, i.e. $\nabla_{\mu} ( T^{\mu\nu} + T_{\mathrm{gauge}}^{\mu\nu}) =0$. For the sake of simplicity and for comparison with the non-covariant discussion, the magnetic field will be neglected; 
in this case Eq. (\ref{VIO1}) leads to the following system of covariant equations:
\begin{eqnarray}
&& {\mathcal E}^{\mu} {\mathcal E}^{\nu} \sigma_{\mu\nu} - \frac{2}{3} \theta {\mathcal E}_{\alpha}{\mathcal E}^{\alpha} - 
\frac{1}{2} u^{\mu} \nabla_{\mu} ( {\mathcal E}_{\alpha} {\mathcal E}^{\alpha}) = 4 \pi j^{\alpha} {\mathcal E}_{\alpha},
\label{VIO1a}\\
&& u^{\mu} \nabla_{\mu} \rho + \theta (\rho+ p) = - {\mathcal E}^{\alpha} j_{\alpha},
\label{VIO1b}
\end{eqnarray}
where $\sigma_{\mu\nu}$ is the shear tensor while $\theta = \nabla_{\mu} u^{\mu}$ is the total expansion; 
both $\theta$ and $\sigma_{\mu\nu}$ appear in the well known covariant decomposition of $\nabla_{\mu} u_{\nu}$ as: 
\begin{equation}
\nabla_{\mu} u_{\nu} = u^{\alpha} (\nabla_{\alpha} u_{\nu}) u_{\nu} + \sigma_{\mu\nu} + \omega_{\nu\mu} + \frac{\theta}{3} {\mathcal P}_{\nu\mu}
\label{VIO1c}
\end{equation}
with ${\mathcal P}_{\nu\mu} = g_{\mu\nu} - u_{\mu} u_{\nu}$; the total relativistic vorticity is antisymmetric for the exchange of the two indices 
(i.e. $\omega_{\nu\mu} = - \omega_{\mu\nu}$) and drops out of Eq. (\ref{VIO1a}) since it is contracted 
with a term which is symmetric for the same exchange. The covariant analog of the first relation
in Eq. (\ref{4two}) can be deduced from Eq. (\ref{5eightA}) and it is given by:
\begin{equation}
j^{\nu} = \frac{1}{4 \pi}  \nabla_{\mu}\biggl[ {\mathcal E}^{\mu} u^{\nu} - {\mathcal E}^{\nu} u^{\mu} \biggr], \qquad {\mathcal B}_{\rho} =0.
\label{VIO1d}
\end{equation}
In the case of a conformally flat geometry $g_{\mu\nu} = a^2(\tau)\eta_{\mu\nu}$, 
  the dictionary between the various parametrizations of the electric fields is 
quite straightforward, namely ${\mathcal E}^{\mu} = (0, \vec{e}/a) = (0, \vec{E}/a^3)$; obviously 
${\mathcal E}_{\mu} = (0, - \vec{e} \,a) = (0, - \vec{E}/a)$. 

The second law of thermodynamics follows by projecting the generally covariant conservation along $u_{\nu}$:
\begin{equation}
u_{\nu} \nabla_{\mu} T^{\mu\nu} = Y^{\nu}_{\,\,\,\,\,\,\alpha}\, j^{\alpha} \, u_{\nu}.
\label{VIO2}
\end{equation}
After having introduced the entropy four-vector  
 ${\mathcal S}^{\mu}$ \cite{seven,eight} Eq. (\ref{VIO2}) implies that 
\begin{equation}
\nabla_{\mu} {\mathcal S}^{\mu} = \frac{Y^{\nu}_{\,\,\,\,\,\,\alpha}\, j^{\alpha} \, u_{\nu}}{T}, \qquad {\mathcal S}^{\mu} = \varsigma \, u^{\mu},
\label{VIO3}
\end{equation}
where, from the fundamental identity of thermodynamics we have $ T\, \varsigma = (\rho + p)$.
In the case of the Ohmic conductor the four-current is given by:
\begin{equation}
j^{\alpha} = \sigma_{\mathrm{c}} Y^{\alpha\beta} u_{\beta} \equiv \sigma_{c} {\mathcal E}^{\alpha},
\label{VIO3a}
\end{equation}
so that the second law of thermodynamics becomes, in this case, 
\begin{equation}
\nabla_{\mu} {\mathcal S}^{\mu} = \frac{8 \pi}{T} \sigma_{c} \rho_{E} \geq 0, \qquad \rho_{E}= - \frac{1}{8\pi} {\mathcal E}_{\alpha} \, {\mathcal E}^{\alpha}.
\label{VIO5}
\end{equation}
Since $\rho_{E}$ is always positive semidefinite we also have that $\nabla_{\mu} {\mathcal S}^{\mu} \geq 0$: this is the covariant 
analog of the Joule heating. In the case of a conformally flat geometry of Friedmann-Robertson-Walker type,
recalling the dictionary discussed after Eq. (\ref{VIO1d}), we have 
that $\rho_{E} = E^2(\tau)/[8\pi a^4(\tau)]$; if $\vec{E}(\tau) = E_{0} a^{\lambda} \hat{n}$  we have, in the case  $\lambda=2$, 
$\rho_{E} = E_{0}^2/(8 \pi)$. Recalling that $ \nabla_{\mu} {\mathcal S}^{\mu} = ( \varsigma^{\prime} + 3 {\mathcal H} \varsigma)/a$, 
 in the non-covariant case Eq. (\ref{VIO5}) 
becomes $\varsigma^{\prime} + 3 {\mathcal H} \varsigma =  8 \pi \sigma \rho_{E}/T$ where, as already stressed
in Eq. (\ref{4zero}),  $\sigma = \sigma_{c} a$ denotes the comoving conductivity.

When the four-current is determined by Eq. (\ref{VIO1d}) the expression of the second law of thermodynamics follows 
from Eq. (\ref{VIO2}) and it can be expressed as:
\begin{equation}
\nabla_{\mu} {\mathcal S}^{\mu} = - \frac{1}{4 \pi T} {\mathcal E}_{\alpha} \nabla_{\beta} \biggl({\mathcal E}^{\beta} u^{\alpha} - {\mathcal E}^{\alpha} u^{\beta} \biggr). 
\label{VIO6}
\end{equation}
After some straightforward algebra the right hand side of Eq. (\ref{VIO6}) can be shown to consist of three terms:  the first one contains
 $(\nabla_{\beta} u^{\beta}) {\mathcal E}_{\alpha} {\mathcal E}^{\alpha}$; the second term is proportional to $ {\mathcal E}_{\alpha} u^{\beta} \nabla_{\beta} {\mathcal E}^{\alpha}$; finally, the third term contains ${\mathcal E}^{\alpha} {\mathcal E}^{\beta} \nabla_{\beta} u_{\alpha}$. 
If the covariant decomposition of Eq. (\ref{VIO1c}) is now inserted into Eq. (\ref{VIO6}) the second law of thermodynamics becomes
\begin{equation}
\nabla_{\mu} {\mathcal S}^{\mu} = - \frac{4\, \theta}{3 \, T} \rho_{E} - \frac{1}{4 \pi T} {\mathcal E}^{\alpha} {\mathcal E}^{\beta} \sigma_{\alpha\beta} - 
\frac{u^{\beta}}{T} \nabla_{\beta} \rho_{E},
\label{VIO7}
\end{equation}
which is nothing but the generally covariant analog of the first equality reported in Eq. (\ref{5tenA}), as it can be easily verified.
Since the shear tensor vanishes when the underlying background is homogeneous and isotropic, Eq. (\ref{VIO7}) only 
depends upon $\rho_{E}$ and its derivative. But whenever $\rho_{E}$ is covariantly 
constant Eq. (\ref{VIO7}) reduces to
\begin{equation}
\nabla_{\mu} {\mathcal S}^{\mu} = - \frac{4\, \theta}{3 \, T} \rho_{E}, \qquad \theta = \nabla_{\mu} u^{\mu}= 3 H,
\label{VIO8}
\end{equation}
and coincides exactly with Eq. (\ref{5tenA}) in the conformally flat case. Equation (\ref{VIO8}) demonstrates that the second law of thermodynamics is only satisfied in the case $H < 0$ (i.e. contracting universe) but not in the expanding case. The general results of Eqs. (\ref{VIO5})--(\ref{VIO8}) also apply when $g_{\mu\nu}(x) = Q(x) \eta_{\mu\nu}$ where $x$ denotes here a generic space-time point. The pair production due to the Schwinger effect has been evaluated in Ref. \cite{five} by studying the Abelian-Higgs model in quasi-de Sitter space, a model already analyzed in the context of inflationary magnetogenesis. In the case of minimally coupled Higgs field the large-scale magnetic fields are minute but they are larger in the non-minimally coupled case (see last paper of Ref. \cite{nine}). Among other things the present analysis demonstrate that the gauge configurations used in Ref. \cite{five} lead to a violation of the second law of thermodynamics when the 
underlying background geometry expands. 

An argument seemingly based on Weyl invariance stipulates that any gauge field configuration is allowed 
in a conformally flat background geometry provided the physical currents 
are appropriately arranged. However, even if the Abelian gauge field equations are 
Weyl invariant in four space-time dimensions, the energy-momentum tensor itself is not. 
Hence the dilution of the electromagnetic energy density in curved backgrounds
cannot be deduced by simply invoking the analogy with the Minkowskian 
space-time. In particular, when the Universe expands the Joule heating 
is always positive semidefinite and the second law of thermodynamics does not depend 
on the sign of the expansion rate.  Conversely the total current needed to maintain
 a constant (electric) energy density in a curved background can be expressed in terms of an effective 
conductivity that depends on the expansion rate and on its overall sign.
The covariant divergence of the entropy four-vector turns out to be
 negative semidefinite when the Universe expand while 
it is positive semidefinite when the Universe contracts. 
This observation ultimately causes the violation 
of the second law  in the case of an expanding background where 
the electric energy density is constant. The conclusions and the bounds obtained from the rate 
of pair production in an expanding de Sitter space-time have been derived in the 
context of a background electric field that violates the second law of thermodynamics.
For this reason they seem neither motivated nor robust.
The violations of the second law are instead absent if and when the 
conformally flat backgrounds contract.

\end{document}